\documentclass{PoS}

\title{Using dipole processes to constrain the flavour of four-fermion effective interactions}

\ShortTitle{Constraining the flavour of $ \psi^4 $ with dipoles}

\author{S.~J{\"a}ger\\
        University of Sussex\\
        E-mail: {\email{S.Jaeger@sussex.ac.uk}}
        }

\author{K.~Leslie\\
        University of Sussex\\
        E-mail: {\email{K.Leslie@sussex.ac.uk}}
        }

\author{\speaker{L.~Vale~Silva} \\
        IFIC, Universitat de Val\`encia - CSIC\\
        E-mail: {\email{luizva@ific.uv.es}}}

\abstract{Dipole interactions encode a rich variety of phenomena, such as radiative decays and electric dipole moments in both quark and lepton sectors, which probe physics beyond the Standard Model up to very high energy scales; due to renormalization, non-dipole operators mix into dipole ones, thus possibly generating observable effects that can be investigated by those same phenomena. I consider four-fermion contact interactions for which the leading order mixing into dipoles happens at two-loops (i.e., when one-loop effects vanish) and for which this mixing can avoid small Yukawa couplings,
and then explore the phenomenological consequences for flavour and $ \mathbf{CP} $ violation coming from New Physics.}

\FullConference{%
  40th International Conference on High Energy physics - ICHEP2020\\
  July 28 - August 6, 2020\\
  Prague, Czech Republic (virtual meeting)
}

\usepackage{slashed}
\usepackage{amsmath}

\usepackage{scalerel,amssymb}

\usepackage{multirow}

\usepackage{graphicx}

\usepackage{caption}
\usepackage{subcaption}

\usepackage{lipsum}
\usepackage{filecontents}
\usepackage[dvipsnames]{xcolor}

\begin{document}

\section{Introduction}

Dipole operators encode a rich variety of phenomena in both quark and lepton sectors, thus testing the Standard Model (SM) structure thoroughly, and probing the flavour structure and amount of $ \mathbf{CP} $ violation of generic extensions of the SM that manifest in dipoles.
Moreover, due to renormalization, non-dipole operators mix into dipole ones, and possibly generate observable effects that can be investigated by the same phenomena that probe directly dipole operators, such as radiative decays and Electric Dipole Moments (EDMs).


Effects from generic extensions of the SM involving new heavy degrees of freedom can be captured in a model-independent way by operators of dimension higher than four involving SM fields only, suppressed by some power of the characteristic scale of New Physics (NP).
The so-called SM Effective Field Theory (SMEFT) framework extends the SM with a complete and minimal set of higher dimensional operators that respect the SM gauged symmetries,
thus providing a universal parametrisation of heavy NP effects invariant under these symmetries.
The one-loop Anomalous Dimension Matrix (ADM) of the full set of dimension-six operators, which is the set of higher dimensional operators on which we focus here, can be found in \cite{Jenkins:2013zja,Jenkins:2013wua,Alonso:2013hga}.
The presence of mixing into dipole operators at one-loop sets important bounds on instances of operators of classes $ H^2 X^2 $ (involving two scalar fields and two field strength tensors), $ X^3 $ (involving three field strength tensors), and $ \psi^4 $ (involving four fermions), see e.g. \cite{Alonso:2013hga,Cirigliano:2019vfc}.

Here, we discuss the mixing into dipoles in some cases where the leading order effect happens at two-loops, i.e., when one-loop ADM elements vanish.
Namely, operators of the type $ \psi^4 $. We then consider phenomenological bounds on the effective coupling of $ \psi^4 $, notably charged lepton radiative decays and EDMs, for which experimental bounds typically probe energy scales much above the direct reach of current and foreseeable colliders.
This extends our preliminary analysis of \cite{ValeSilva:2019qcf}, in which $ \psi^2 H^3 $ operators (involving two fermions and three scalars) were discussed.

\section{Unsuppressed mixing effects}\label{sec:basis}

We consider the SM Lagrangian $ \mathcal{L}_{\rm SM} $, to which we add right-handed neutrinos to enlarge the scope of our analysis, and dimension-six operators, $ \mathcal{L} \equiv \mathcal{L}_{{\rm SM} + \nu_R} + \sum_i C_i Q_i $.
A basis of dimension-six operators $ Q_i $ can be found in \cite{Buchmuller:1985jz, Grzadkowski:2010es}, hereafter called the Warsaw basis, while a basis of operators of dimension-six involving right-handed neutrinos can be found in, e.g., \cite{Liao:2016qyd}.

\begin{figure}[t]
	\centering
	\includegraphics[scale=0.25]{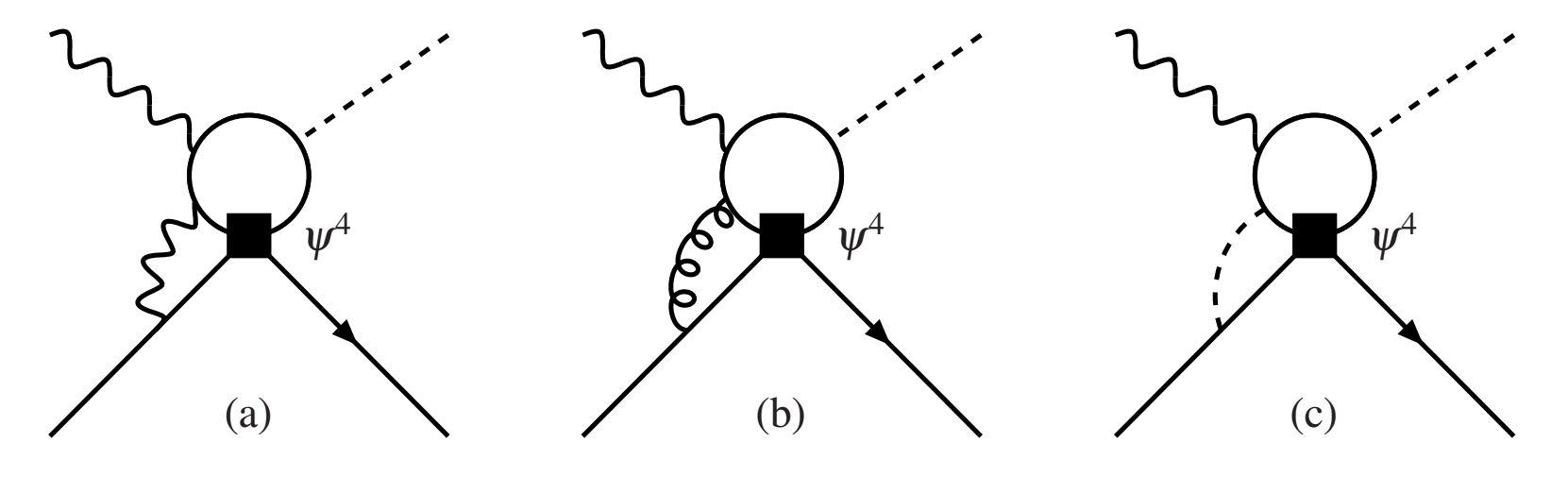}
	\caption{Sample of two-loop diagrams required in the determination of the mixing of four-fermion operators $ \psi^4 $ (whose interaction vertex insertions are represented by a filled box) into dipole ones; diagram (a) shows the exchange of an EW gauge boson, diagram (b) shows the exchange of a gluon, diagram (c) shows the exchange of a scalar. Diagram (b) stresses effects generated by QCD.}\label{fig:fig1}
\end{figure}

Four-fermion operators are divided into 5 categories in the Warsaw basis, according to the chiralities of the fields involved, schematically:\footnote{The notation $ L $ ($ R $) designates left-handed (respec., right-handed) Dirac fields; the Dirac structure is omitted.} $ (\overline{L} L) (\overline{L} L) $, $ (\overline{R} R) (\overline{R} R) $, $ (\overline{L} L) (\overline{R} R) $ (that we further subdivide into semi-leptonic cases, SL, and purely leptonic and four-quark cases, 4L and 4Q, respectively), $ (\overline{L} R) (\overline{R} L) $ and $ (\overline{L} R) (\overline{L} R) $ (together with their Hermitian conjugates in the latter two cases). We restrict the scope of the operators considered here due to phenomenological reasons, focusing at the moment on those for which the mixing into dipoles can avoid small Yukawa couplings in cases where the dipole process involves light external flavours: diagrammatically, operators of the categories 4L-$ (\overline{L} L) (\overline{R} R) $, 4Q-$ (\overline{L} L) (\overline{R} R) $, $ (\overline{L} R) (\overline{R} L) $ and $ (\overline{L} R) (\overline{L} R) $ can lead to diagrams (a) and (b) in Fig.~\ref{fig:fig1}; instead, operators of the categories $ (\overline{X} X) (\overline{X} X) $, $ X = L, R $, and SL-$ (\overline{L} L) (\overline{R} R) $ necessarily require attaching scalar fields to the external fermion lines to build diagrams that contribute to Green's functions involving two fermions, one scalar and one gauge boson; this thus leads to an overall Yukawa suppression if the external flavours are light, an example of a diagram being given by (c) in Fig.~\ref{fig:fig1}.
Among the operators that possibly avoid Yukawa suppression, we have operators that mix into dipoles already at one-loop order: the operators\footnote{The fields $ \ell $ and $ q $ ($ e $, $ u $ and $ d $) are doublets (respec., singlets) under the gauge symmetry $ SU (2)_L $ whose indices are $ j, k $; $ p, r, s, t $ are generation indices; $ \varepsilon $ is the anti-symmetric symbol involving two indices.}
$ Q^{(1), p r s t}_{\ell e q u} \equiv (\bar{\ell}^j_p e_r) \varepsilon_{j k} (\bar{q}^k_s u_t) $
mix into
$ Q^{(3), p r s t}_{\ell e q u} \equiv (\bar{\ell}^j_p \sigma_{\mu \nu} e_r) \varepsilon_{j k} (\bar{q}^k_s \sigma^{\mu \nu} u_t) $
at one-loop, while the latter operators mix into dipoles at one-loop (the same comment would also apply to operators similarly defined involving instead right-handed neutrinos and down-type quarks).
Therefore, we focus on the following operators ($ T^A $ are Gell-Mann matrices)

\begin{eqnarray}
	& Q_{\ell e d q}^{p r s t} \equiv (\bar{\ell}_p^j e_t) (\bar{d}_s q_{r, j}) \,, \quad & Q_{\ell e}^{p r s t} \equiv (\bar{\ell}_p \gamma_\mu \ell_r) (\bar{e}_s \gamma^\mu e_t) \,, \nonumber\\
	& Q^{(1), p r s t}_{q \xi} \equiv (\bar{q}_p \gamma_\mu q_r) (\bar{\xi}_s \gamma^\mu \xi_t) \,, \quad & Q^{(8), p r s t}_{q \xi} \equiv (\bar{q}_p \gamma_\mu T^A q_r) (\bar{\xi}_s \gamma^\mu T^A \xi_t) \,, \quad \xi = u, d \,, \nonumber\\
	& Q^{(1), p r s t}_{q u q d} \equiv (\bar{q}^j_p u_r) \varepsilon_{jk} (\bar{q}^k_s d_t) \,, \quad & Q^{(8), p r s t}_{q u q d} \equiv (\bar{q}^j_p T^A u_r) \varepsilon_{jk} (\bar{q}^k_s T^A d_t) \,, \nonumber
\end{eqnarray}
whose leading order mixing into dipoles happens at two-loops.
This set of operators can easily be enlarged to include right-handed neutrinos.
Other than possibly involving large Yukawas as mentioned previously, the mixing of four-fermion operators into dipoles can involve strong couplings (compare diagrams (a) and (b) in Fig.~\ref{fig:fig1}), and moreover can be enhanced by large (color) group factors. This impacts the phenomenology of the respective operators, since it leads to stronger bounds on the corresponding Wilson coefficients (WCs). A similar discussion can be found in the SM in the context of $ b \to s \gamma $ transitions, see e.g. \cite{Bertolini:1986th,Deshpande:1987nr}.

Under Fierz transformations the operators $ Q_{\ell e d q} $, $ Q_{\ell e} $, $ Q^{(1)}_{q \xi} $, $ Q^{(8)}_{q \xi} $, $ \xi = u, d $, have the same chiral structure, $ (\overline{L} R) (\overline{R} L) $. Diagrams (a) and (b) in Fig.~\ref{fig:fig1} with insertions of operators of such chiral structure would typically result from Barr-Zee diagrams \cite{Barr:1990vd} where a heavy scalar field (of mass $ \sim \Lambda $ much above the EW scale) is exchanged.

\section{Phenomenology}\label{sec:GreensFunctions}



Hereafter, we discuss some phenomenological bounds on the WCs of four-fermion operators; a complete analysis will be provided in a future publication. Useful expressions and comments, in particular concerning the extraction of the renormalization constants, can already be found in \cite{ValeSilva:2019qcf}.


A sample of bounds is summarized in Tab.~\ref{tab:tab1}.
They do not include yet renormalization effects below the EW scale, such as the mixing of four-fermion operators into dipoles below the EW scale (that arrives at two-loops as well for the cases considered here). Therefore, they are preliminary estimates that result uniquely from the leading order mixing at two-loops of four-fermion operators into dipoles in SMEFT. A future publication will also discuss hadronic uncertainties.
We now comment in turn on the bounds provided in Tab.~\ref{tab:tab1}:


\vspace{2mm}
\textbf{$ \mathbf{CP} $ violation in quark dipoles.} NP (and SM) sources of $ \mathbf{CP} $ violation can for instance be probed by experimental bounds on neutron and atomic EDMs, which are sensitive to $ \mathbf{CP} $ violation in quark electric and chromo-electric dipoles, among other effective operators, see e.g. \cite{Pospelov:2005pr}.
Operators $ Q^{(1)}_{q u} $, $ Q^{(8)}_{q u} $, $ Q^{(1)}_{q u q d} $ and $ Q^{(8)}_{q u q d} $ generate quark dipoles at two-loops which can possibly involve the Yukawa of the top, while operators $ Q^{(1)}_{q d} $, $ Q^{(8)}_{q d} $, $ Q_{\ell e d q} $ may lead to contributions at two-loops that involve the Yukawas of the bottom or the tau (among other large Yukawas compared to the down- or up-quark external/valence flavours).
In the case of these four-quark operators involving the top, there is no analogous effect of four-fermion mixing into dipoles at one- or two-loops in a new effective field theory defined much below the EW scale, where the top is integrated out together with the heavy gauge bosons and the Higgs scalar.
Although induced at two-loops, powerful bounds follow from the many available unsuppressed factors: top Yukawa, strong coupling and color factor enhancement.
We provide in the topmost part of Tab.~\ref{tab:tab1} bounds on the WCs of the operators $ Q^{(1)}_{q u} $ and $ Q^{(8)}_{q u} $ involving the top, which are constrained to be smaller than $ \lesssim ( 700 \; \text{TeV} )^{-2} $ to $ ( 3000 \; \text{TeV} )^{-2} $. These bounds originate from chromo-electric dipole contributions to Hg-EDM
(leading to an effective $ \mathbf{CP} $ violating pion-nucleon-nucleon coupling, see e.g. \cite{Pospelov:2005pr}), while the bounds extracted from neutron-EDM \cite{Abel:2020gbr} and resulting from electric dipole contributions are weaker by a factor $ \mathcal{O} (20) - \mathcal{O} (50) $.


\vspace{2mm}
\textbf{Charged lepton dipoles.} Powerful bounds can also be set on NP contributions to leptonic dipoles when sources of $ \mathbf{CP} $ violation and Lepton Flavour Violation (LFV) are present. The operators $ Q_{\ell e} $ and $ Q_{\ell e d q} $ can lead to mixing into dipole operators proportional to tau and bottom Yukawa couplings, respectively, much larger compared to the light external electron and muon flavours in $ e $-EDM and $ \mu \to e \gamma $ transition. Note that below the EW scale four-fermion operators involving the bottom or the tau can mix into dipoles at two-loops (not at one-loop order for the operators under consideration here); the corresponding ADM elements that involve EM couplings have not been determined yet (an estimate of their phenomenological impact is given in \cite{Crivellin:2017rmk}). Some bounds on the NP WCs are provided in Tab.~\ref{tab:tab1}, showing that one can probe energy scales as large as $ \sim 10 \; \text{TeV} $ to $ 400 \; \text{TeV} $.
In the case of $ e $-EDM, similar bounds have been found by \cite{Panico:2018hal}.


As discussed in \cite{Crivellin:2017rmk}, the operator $ Q_{\ell e d q} $ generates one-loop contributions to $ \mu \to e $ conversion in nuclei \cite{Bertl:2006up} through the operator $ m_\mu (\bar{e} P_X \mu) G^A_{\nu \rho} G^{\nu \rho}_A $, $ X = L, R $, of dimension higher than six (Ref.~\cite{Crivellin:2017rmk} discusses NP phenomena below the EW scale; $ G^A_{\nu \rho} $ is the field-strength tensor for gluons), leading in the case of $ Q_{\ell e d q} $ to a bound on its Wilson coefficient stronger by a factor of a few compared to the one derived from the radiative decay $ \mu \to e \gamma $, shown in Tab.~\ref{tab:tab1}.


\vspace{2mm}
Throughout the previous discussion, other heavy flavours, such as the charm (much heavier than the external flavours of the processes studied above), could also be considered.



\begin{table}
\centering
\renewcommand{\arraystretch}{1.6}
\begin{tabular}{c|ccc}
\hline
& Observable & Couplings & Bound \\
\hline
\hline
$ Q^{(1)}_{qu} $ & \multirow{2}{*}{Hg-EDM \cite{Graner:2016ses}} & $ y_{t} \times | {\rm Im} [ \tilde{C}^{(1), uttu}_{qu} (\Lambda) ] | $ & {\color{black} $ \lesssim \mathcal{O} (10^{-6}) \, {\rm TeV}^{-2} $ } \\
$ Q^{(8)}_{qu} $ & & $ y_{t} \times | {\rm Im} [ \tilde{C}^{(8), uttu}_{qu} (\Lambda) ] | $ & {\color{black} $ \lesssim \mathcal{O} (10^{-7}) \, {\rm TeV}^{-2} $ } \\
\hline
\hline
\multirow{2}{*}{$ Q_{\ell e} $} & $ \mathcal{B} ( \mu \to e \gamma ) $ \cite{TheMEG:2016wtm} & $ y_{\tau} \times \sqrt{ | \tilde{C}^{e \tau \tau \mu}_{\ell e} (\Lambda) |^2 + | \tilde{C}^{\mu \tau \tau e}_{\ell e} (\Lambda) |^2 } $ & {\color{black} $ \lesssim \mathcal{O}(10^{-5}) \; {\rm TeV}^{-2} $ } \\
& $ e $-EDM \cite{Andreev:2018ayy} & $ y_{\tau} \times | {\rm Im} [ \tilde{C}^{e \tau \tau e}_{\ell e} (\Lambda) ] | $ & {\color{black} $ \lesssim \mathcal{O} (10^{-7}) \, {\rm TeV}^{-2} $ } \\
\hline
\hline
\multirow{2}{*}{$ Q_{\ell e d q} $} & $ \mathcal{B} ( \mu \to e \gamma ) $ \cite{TheMEG:2016wtm} & $ y_{b} \times \sqrt{ | \tilde{C}^{e b b \mu}_{\ell e d q} (\Lambda) |^2 + | \tilde{C}^{\mu b b e}_{\ell e d q} (\Lambda) |^2 } $ & {\color{black} $ \lesssim \mathcal{O}(10^{-5}) \; {\rm TeV}^{-2} $ } \\
& $ e $-EDM \cite{Andreev:2018ayy} & $ y_{b} \times | {\rm Im} [ \tilde{C}^{e b b e}_{\ell e d q} (\Lambda) ] | $ & {\color{black} $ \lesssim \mathcal{O} (10^{-7}) \, {\rm TeV}^{-2} $ } \\
\hline
\end{tabular}
\caption{Some preliminary bounds on the WCs $ C $ of the operators indicated in the first column. The considered observables are those in the second column. We indicate in the third column the combination of WCs at the NP scale $ \Lambda $ (tildes are used to indicate that we move from the interaction to the mass basis), where flavour indices are shown, with the types of Yukawa factors ($ y_t $, $ y_b $, $ y_\tau $ stand for top, bottom, and tau Yukawas, respectively), to which the bounds in the fourth column apply. See text for more comments.}\label{tab:tab1}
\end{table}


\section{Conclusions}\label{sec:discussion}

We have discussed how operators of the class $ \psi^4 $ can be probed indirectly by their mixing-induced contributions to dipole operators, resulting in a broad set of phenomenological applications involving both lepton and quark sectors.
The leading order mixing of $ \psi^4 $ operators into dipoles at two-loops leads to important bounds on the Wilson coefficient of $ \psi^4 $ derived from $ \mathcal{B} (\mu \to e \gamma) $ and EDMs, thus showing the power of dipoles in probing the flavour structure of NP, including the one encoded in non-dipole operators.
The dimension-six operators $ \psi^4 $ are generated in many extensions of the SM involving new heavy degrees of freedom, that can therefore be constrained by the bounds discussed previously.

We stress the fact that such two-loop contributions can be proportional to large Yukawa couplings in processes involving light external flavours, and/or color enhanced. The set of renormalization constants describing the mixing of physical four-fermion operators into dipoles, together with a complete phenomenological analysis, is under preparation, where operators of the category $ \psi^2 H^3 $, see \cite{ValeSilva:2019qcf}, will also be discussed.

\newpage{\pagestyle{empty}\cleardoublepage}


\vspace*{3mm}

\begin{thebibliography}{100}


\bibitem{Jenkins:2013zja} E.~E.~Jenkins, A.~V.~Manohar and M.~Trott,
  JHEP {\bf 1310}, 087 (2013)
  doi:10.1007/JHEP10(2013)087
  [arXiv:1308.2627 [hep-ph]].

\bibitem{Jenkins:2013wua} E.~E.~Jenkins, A.~V.~Manohar and M.~Trott,
  JHEP {\bf 1401}, 035 (2014)
  doi:10.1007/JHEP01(2014)035
  [arXiv:1310.4838 [hep-ph]].

\bibitem{Alonso:2013hga} R.~Alonso, E.~E.~Jenkins, A.~V.~Manohar and M.~Trott,
  JHEP {\bf 1404}, 159 (2014)
  doi:10.1007/JHEP04(2014)159
  [arXiv:1312.2014 [hep-ph]].

\bibitem{Cirigliano:2019vfc} V.~Cirigliano, A.~Crivellin, W.~Dekens, J.~de Vries, M.~Hoferichter and E.~Mereghetti,
  Phys.\ Rev.\ Lett.\  {\bf 123}, no. 5, 051801 (2019)
  doi:10.1103/PhysRevLett.123.051801
  [arXiv:1903.03625 [hep-ph]].


\bibitem{ValeSilva:2019qcf}
L.~Vale Silva, S.~Jaeger and K.~Leslie,
PoS \textbf{EPS-HEP2019}, 248 (2020)
doi:10.22323/1.364.0248

\bibitem{Buchmuller:1985jz}
W.~Buchmuller and D.~Wyler,
Nucl. Phys. B \textbf{268}, 621-653 (1986)
doi:10.1016/0550-3213(86)90262-2

\bibitem{Grzadkowski:2010es} B.~Grzadkowski, M.~Iskrzynski, M.~Misiak and J.~Rosiek,
  JHEP {\bf 1010}, 085 (2010)
  doi:10.1007/JHEP10(2010)085
  [arXiv:1008.4884 [hep-ph]].

\bibitem{Liao:2016qyd} Y.~Liao and X.~D.~Ma,
  Phys.\ Rev.\ D {\bf 96}, no. 1, 015012 (2017)
  doi:10.1103/PhysRevD.96.015012
  [arXiv:1612.04527 [hep-ph]].

\bibitem{Bertolini:1986th} S.~Bertolini, F.~Borzumati and A.~Masiero,
Phys. Rev. Lett. \textbf{59}, 180 (1987)
doi:10.1103/PhysRevLett.59.180

\bibitem{Deshpande:1987nr} N.~G.~Deshpande, P.~Lo, J.~Trampetic, G.~Eilam and P.~Singer,
Phys. Rev. Lett. \textbf{59}, 183-185 (1987)
doi:10.1103/PhysRevLett.59.183

\bibitem{Barr:1990vd}
S.~M.~Barr and A.~Zee,
Phys. Rev. Lett. \textbf{65}, 21-24 (1990)
[erratum: Phys. Rev. Lett. \textbf{65}, 2920 (1990)]
doi:10.1103/PhysRevLett.65.21


\bibitem{Pospelov:2005pr} M.~Pospelov and A.~Ritz,
  Annals Phys.\  {\bf 318}, 119 (2005)
  doi:10.1016/j.aop.2005.04.002
  [hep-ph/0504231].

\bibitem{Abel:2020gbr}
C.~Abel \textit{et al.} [nEDM],
Phys. Rev. Lett. \textbf{124}, no.8, 081803 (2020)
doi:10.1103/PhysRevLett.124.081803
[arXiv:2001.11966 [hep-ex]].

\bibitem{Crivellin:2017rmk} A.~Crivellin, S.~Davidson, G.~M.~Pruna and A.~Signer,
  JHEP {\bf 1705}, 117 (2017)
  doi:10.1007/JHEP05(2017)117
  [arXiv:1702.03020 [hep-ph]].

\bibitem{Panico:2018hal} G.~Panico, A.~Pomarol and M.~Riembau,
  JHEP {\bf 1904}, 090 (2019)
  doi:10.1007/JHEP04(2019)090
  [arXiv:1810.09413 [hep-ph]].

\bibitem{Bertl:2006up}
W.~H.~Bertl \textit{et al.} [SINDRUM II],
Eur. Phys. J. C \textbf{47}, 337-346 (2006)
doi:10.1140/epjc/s2006-02582-x

\bibitem{Graner:2016ses} B.~Graner, Y.~Chen, E.~G.~Lindahl and B.~R.~Heckel,
Phys. Rev. Lett. \textbf{116}, no.16, 161601 (2016)
[erratum: Phys. Rev. Lett. \textbf{119}, no.11, 119901 (2017)]
doi:10.1103/PhysRevLett.116.161601
[arXiv:1601.04339 [physics.atom-ph]].

\bibitem{TheMEG:2016wtm} A.~M.~Baldini {\it et al.} [MEG Collaboration],
  Eur.\ Phys.\ J.\ C {\bf 76}, no. 8, 434 (2016)
  doi:10.1140/epjc/s10052-016-4271-x
  [arXiv:1605.05081 [hep-ex]].

\bibitem{Andreev:2018ayy} V.~Andreev {\it et al.} [ACME Collaboration],
  Nature {\bf 562}, no. 7727, 355 (2018).
  doi:10.1038/s41586-018-0599-8


\end{thebibliography}
\end{document}